\documentclass[12pt]{article}
\usepackage{amsfonts,amsmath,amsxtra}
\usepackage{amssymb}
\usepackage[utf8]{inputenc}
\usepackage[normalem]{ulem}
\usepackage[russian,english]{babel}
\usepackage{xcolor}
\usepackage{pst-circ}
\def\Id{\operatorname{Id}}

\def\C{\mathbb{C}}

\def\beq{\begin{equation}}
	\def\eeq{\end{equation}}
\def\beqq{\begin{equation*}}
	\def\eeqq{\end{equation*}}
\def\bl{\boldsymbol{\lambda}}

\def\brho{\boldsymbol{\rho}}
\def\d{\partial}

\def\Im{\operatorname{Im}}
\def\g{\gamma}
\def\gl{\mathfrak{gl}}

\def\Id{\operatorname{Id}}

\def\jj{{\mathbf k}}
\def\kk{{\mathbf k}}

\def\l{\lambda}

\def\R{\mathbb{R}}
\def\Z{\mathbb{Z}}

\def\Re{\mathrm{Re}\,}

\def\T{T^{g}}
\def\ts{\textstyle}

\def\Z{\mathbb{Z}}

\def\ws{\hfill{$\square$}}
\def\bs{\begin{subequations}\label}
\def\es{\end{subequations}}
\newcommand{\ba}{\begin{align}}
\newcommand{\ea}{\end{align}}

\newtheorem{lemma}{Lemma}[section]
\newtheorem{proposition}{Proposition}[section]

\newtheorem{theorem}{Theorem}

\newcommand{\rf}[1]{(\ref{#1})}
\begin{document}

\begin{center}
\phantom{1}

\vspace{-2cm}
\hfill ITEP-TH-20/21\\
\hfill IITP-TH-15/21\\ [20mm]
		{\bf \large Wave function for  $GL(n,\mathbb{R})$ hyperbolic Sutherland model}
		\bigskip
		
		{\bf S. Kharchev$^{\,\star\,\natural}$,\, S. Khoroshkin$^{\,\star\,\circ}$,
		}\medskip\\
		$^\star${\it Institute for Theoretical and Experimental Physics, B. Cheremushkinskaya, 25, Moscow 117259, Russia;}\smallskip\\
		$^\natural${\it
			Institute for Information Transmission Problems RAS (Kharkevich Institute),Bolshoy Karetny per. 19, Moscow, 127994, Russia;}\smallskip\\
		$^\circ${\it National Research University Higher School of Economics, Moscow, Russia.}
	\end{center}
	\begin{abstract}\noindent
We obtain certain Mellin-Barnes integrals which present wave functions for  $GL(n,\R)$ hyperbolic Sutherland model with arbitrary positive coupling constant.
	\end{abstract}	

%\thispagestyle{empty}
%\footnotesize
%\tableofcontents
%\normalsize
%\newpage
%\clearpage
\setcounter{page}{1}

\section{Introduction}
In the paper \cite{GKL} A. Gerasimov, S. Kharchev and D. Lebedev applied the famous technique of Gelfand--Zetlin basis \cite{GT} for the derivation of integral presentation  for $GL(n,\R)$ Whittaker functions, equivalently, for  wave functions of the open Toda chain. They used formulas for the action of Lie algebra generators on Gelfand--Zetlin patterns  to construct certain infinite-dimensional representation of the Lie algebra $\gl(n,\R)$ in the functional space of meromorphic functions, where the Lie algebra acts by difference operators with rational coefficients. In this representation the Whittaker vectors and  nondegenerate pairing where found, so that the pairing of two dual Whittaker vectors gives Mellin-Barnes presentation
 for  the Whittaker functions. The  mentioned Whittaker vectors are given by products of Euler Gamma functions, and the pairing
 is the integration on the imaginary plane in $\C^{n(n-1)/2}$ with the
  Sklyanin measure also factorized into a product of Gamma functions
	
Besides, in \cite{GKL} the Mellin-Barnes presentation for zonal spherical functions of the symmetric space  $GL(n,\R)/O(n)$ was obtained, which are, in  turn, the wave functions of the hyperbolic Sutherland model for a special value of the coupling constant.
Precisely, in the above infinite dimensional representation of $\gl(n)$ the spherical vector was found. It is given by another products of Euler Gamma functions, so that the corresponding matrix elements is an eigenvector of Sutherland operator
\beq\label{i2}
H_2=-\sum_{i=1}^n\frac{\d^2}{\d x_i^2}+2\sum_{i<j}\frac{g(g-1)}{\sh^2(x_i-x_j)},
\eeq	
again presented by the integral of Mellin-Barnes type for the particular case $g=1/2$.

Thus in the framework of Representation Theory, the wave function for Sutherland model with the coupling constant $g=1/2$ admits the integral presentation
\beq\label{i1}
\begin{split}
\Psi^{(1/2)}_{\l_1,\ldots, \l_n}&(x_1,\ldots,x_n)=\prod_{i<j}\sh^{1/2}|x_i-x_j|\times\\ \int\limits_{i\R^{\frac{n(n-1)}{2}}}
\prod\limits_{i=1}^{n-1}&\frac{\prod\limits_{j=1}^{i}
\prod\limits_{k=1}^{i+1}\Gamma\Big(\frac{\g_{i,j}-\g_{i+1,k}}{2}+\frac{1}{4}\Big)
\Gamma\left(\frac{\g_{i+1,k}-\g_{i,j}}{2}+\frac{1}{4}\right)}
{\prod\limits_{1\leq r\neq s\leq i}\Gamma\left(\g_{i,r}-\g_{i,s}\right)}
			e^{\sum_{i,j=1}^n(\g_{i,j}-\g_{i-1,j})x_i}
		\prod\limits_{\stackrel{i=1}{j\leq i}}^{n-1}d\g_{i,j},
			\end{split}
	\eeq
where $\l_i=\g_{n,i}$, $\l_i\in\imath\R$ and $\g_{i,j}=0$ if $i<j$.

 On the other hand, in their research on generalized hypergeometric functions associated to root systems, G. Heckman and E. Opdam studied in particular the properties of the wave functions of Sutherland Hamiltonian for general coupling constant $g$. G. Heckman  showed in \cite{H} the existence of  the wave function $\Psi^{(g)}_{\l_1,\ldots, \l_n}(x_1,\ldots,x_n)$ such that the function
		\beq \label{phi} \Phi^{(g)}_{\l_1,\ldots, \l_n}(x_1,\ldots,x_n)=\prod_{i<j}\sh^{-g}|x_i-x_j| \ \Psi^{(g)}_{\l_1,\ldots, \l_n}(x_1,\ldots,x_n)\eeq
	 (now called Heckman-Opdam hypergeometric function) is real analytical and invariant with respect to the permutations of the coordinates $x_k$. See also \cite{O} for their further analytical  properties. These results were obtained  by studying the recurrence relations on the coefficients of Taylor expansions of the solutions to corresponding differential equation.

 Our paper is devoted to precise construction of the Sutherland wave function.
This results to analytical version of Heckman-Opdam hypergeometric series
in the case of $GL(n,\R)$. Set
\beq\label{i3}
\begin{split} \Psi^{(g)}_{\l_1,\ldots, \l_n}&(x_1,\ldots,x_n)
=\prod_{j<k}\sh^{g}|x_j-x_k|\times\\
 \int\limits_{\imath\R^{\frac{n(n-1)}{2}}}\prod\limits_{i=1}^{n-1}&
 \frac{\prod\limits_{j=1}^{i}\prod\limits_{k=1}^{i+1}
 \Gamma\left(\frac{\g_{i,j}-\g_{i+1,k}+g}{2}\right) \Gamma\left(\frac{\g_{i+1,k}-\g_{i,j}+g}{2}\right)}
 {\prod\limits_{1\leq r\neq s\leq i}\Gamma\left(\frac{\g_{i,r}-\g_{i,s}}{2}\right)
 \Gamma\left(\frac{\g_{i,r}-\g_{i,s}+2g}{2}\right)}
	 	e^{\sum_{i,j=1}^n(\g_{i,j}-\g_{i-1,j})x_i}
	 	\prod\limits_{\stackrel{i=1}{j\leq i}}^{n-1}d\g_{i,j}
	 \end{split}
	 \eeq	
	 where $\l_i=\g_{n,i}$, $\l_i\in \imath\R$, $g>0$, and $\g_{i,j}=0$ if $i<j$.
	  Our main result is
	 \begin{theorem}\label{th1}
The  function   $\Psi^{(g)}_{\l_1,\ldots, \l_n}(x_1,\ldots,x_n)$
	 		is the eigenfunction of the Hamiltonian \rf{i2},
	 	$$H_2\Psi^{(g)}_{\l_1,\ldots, \l_n}(x_1,\ldots,x_n)=-\bl^2\,
\Psi^{(g)}_{\l_1,\ldots, \l_n}(x_1,\ldots,x_n).$$
{ where $\bl^2=\l_1^2+\ldots+\l_n^2$.}	
 	\end{theorem}
Due to the strong convergency  of the integral, see \rf{iorgov}  the hypergeometric function %\\ $\Phi^{(g)}_{\l_1,\ldots, \l_n}(x_1,\ldots,x_n)$,
	  \rf{phi} is analytical in $(x_1,\ldots,x_n)$ in
a small strip around real hyperplane (see \rf{strip} for more detail).
%a vicinity of $\R^n\subset\C^n$ (see \rf{strip} ).

The crucial new point in the formula \rf{i3} is the denominator of the integral kernel, which can be regarded as a deformation of Sklyanin measure,  or { as} a degeneration of the weight function used in the scalar product of Macdonald polynomials \cite{M}. It may reveals a new type of Barnes integrals associated to integrable systems related to DAHA \cite{Ch} and quantum toroidal algebras \cite{FJM}.
 	
 	Our proof of Theorem \ref{th1} is essentially simple, but uses unexpected arguments from the representation theory. Usually, the technique of matrix elements in the group theory works for special values of parameters, related to real, complex of quaternionic symmetric spaces. The formula \rf{i1} is a typical example of this approach, yielding the wave function only for $g=1/2$. However, we use Laplace operator in Gelfand--Zetlin representation of Lie algebra
 	$\gl(n,\R)$ constructed in \cite{GKL} as a Hamiltonian in the space of rational functions and derive, using its properties, the second order differential equation on the integral in the right hand site of \rf{i3} for arbitrary positive coupling constant. The coupling constant $g$ appears as a parameter $\imath\hbar/2$ in the representation, constructed in \cite{GKL}.

 		Note also that we  use Gelfand--Zetlin formulas in a full range, for all generators of Lie algebra $\gl(n,\R)$.
 	
 	The plan of the paper is as follows. In Section 2, following \cite{GKL}, we collect necessary information about the so called Gelfand--Zetlin representation. We are interested in the second order Laplace operator and { in rational functions $a_{i,j}^{\jj}(\g)$ and
 	$b_{i,j}^{\jj}(\g)$, which constitute constant terms of second order difference operators $e_{i,j}e_{j,i}$. Here $e_{k,l}$ are standard generators  of the Lie algebra $\gl(n,\R)$. For this purpose} we present (well known to specialists) precise expressions for { the action of all generators $e_{k,l}$}  and formulate basic identities on { the mentioned} rational functions responsible for the validness of $\gl(n,\R)$ commutation relations. Here the shifts of the arguments by $2g$ are essential.

In Section 3 we derive the differential equation on the wave function given by integral \rf{i3}. To do this, we imitate standard tricks with Laplace operator, where the rational coefficients $a_{i,j}^{\jj}(\g)$ and $b_{i,j}^{\jj}(\g)$ are used now for a number of proper deformations of the integration contour. Take note here on Lemma \ref{lemma4}, which establishes the difference relations on these coefficients and the integration kernel with the shift by step $2$. It indicates the use of some two--periodic properties of Gelfand--Zetlin coefficients.

Finally, in Section 4 we give $n=2$ example, which can be regarded as Barnes integral presentation of the Legendre function.

\medskip
After this work was completed, we found that
M. Halln\" as and S. Ruijsenaars wrote a series of papers \cite{HR1, HR2, HR3}, where they suggest   a general construction of eigenfunctions for Ruijsenaars systems based on a precise kernel function found in \cite{R}. In particular, degeneration of their construction to hyperbolic Sutherland system \cite{HR2} yields another presentation of the wave functions given by iterated beta integrals over space variables.

 	 \setcounter{equation}{0}
 	 \section{ Gelfand-Zetlin representation}
 	 \subsection{Laplace operator}
 	 In the paper \cite{GKL} A. Gerasimov, S. Kharchev and D. Lebedev used famous	Gelfand--Zetlin basis for the construction of infinite--dimensional representation of the Lie algebra $\gl(n)$, which we also name as Gelfand--Zetlin representation. More precisely, they interpreted Gelfand--Zetlin formulas \cite{GT}  for the action of simple root generators in finite-dimensional representations of $\gl(n,\R)$ as difference operators, presenting the action of $\gl(n,\R)$ in the space of meromorphic functions of $n(n-1)/2$ variables.
 	
 	 Rewrite formulas \cite[(2.1)]{GKL} replacing $\imath\hbar$ factor by parameter $2g$:
\bs{gz1}
\ba
&e_{i,i}=\frac{1}{2g}
\Big(\sum_{k=1}^i\g_{i,k}-\sum_{k=1}^{i-1}\g_{i-1,k}\Big),\label{gz0a}\\
&e_{i,i+1}=-\frac{1}{2g}\sum_{k=1}^{i}
\frac{\prod_{r=1}^{i+1}(\g_{i,k}-\g_{i+1,r}-g)}{\prod_{s\not=k}(\g_{i,k}-\g_{i,s})}\,
T_{\g_{i,k}}^{-g},\label{gz0b}\\
&e_{i+1,i}=\frac{1}{2g}\sum_{k=1}^{i}
\frac{\prod_{r=1}^{i-1}(\g_{i,k}-\g_{i-1,r}+g)}{\prod_{s\not=k}
(\g_{i,k}-\g_{i,s})}\,T_{\g_{i,k}}^{g}.\label{gz0c}		
\end{align}
\es
 	 Here $T_\g^g f(\g)=f(\g+2g)$. If $i$ ranges from $1$ to $n$ and $\g_{n,i}$ specialize to constants $\g_{n,i}=\l_i$, $i=1,\ldots,n$, the
 	 relations \rf{gz1} define
 	 %a homomorphism  of the universal enveloping algebra
 	% $U(\gl(n,\R)) $ to the algebra of difference operators on the variables $\g_{i,j}$,
% $1\leq j\leq i<n$ with rational coefficients, that is
  a representation of  $U(\gl(n,\R))$ in the space of meromorphic  functions on  $\g_{i,j}$, $1\leq j\leq i<n-1$, realized by difference operators with  the step $2g$ and some rational coefficients. It was proved in \cite{GKL} that the center $\operatorname{Z}(\gl(n,\R))$ of the algebra $U(\gl(n,\R))$ acts by scalar operators.

   We need here  precise evaluation of the first two Laplace operators,
 	  \beqq
 L_1=\sum_{i=1}^n e_{i,i},\qquad L_2=\sum_{i,j=1}^n e_{i,j}e_{j,i}.
 \eeqq
 	  \begin{lemma}\label{lemma1} Laplace operators  $L_1$ and $L_2$ are realized in Gelfand-Zetlin representation by the following scalar operators
 	 \beq\label{gz2}
 L_1=\frac{1}{2g}\Big(\sum _j\l_j\Big)\Id,\ \ \ \ L_2=\frac{1}{4g^2}\left(\bl^2-4g^2\brho^2\right)\Id,
 	 \eeq
 where $\brho$ is the Weyl vector with components $\rho_i=\frac{1}{2}(n-2i+1)$,
  $(i=1,\ldots,n)$ such that $\brho^2=\frac{n(n^2-1)}{12}$.
 	 \end{lemma}
 	 {\bf Proof}. The relations \rf{gz2} can be extracted from the relation \cite[(2.19)]{GKL} where the generating function of central elements in Gelfand-Zetlin representation was computed in a form of Capelli determinant. \ws

\subsection{Root vectors and related identities}

We need further formulas for the action of all root vectors $e_{i,j}$ in Gelfand--Zetlin representation.
Using the recurrent relations
\bs{a3}
\ba
&e_{i,i+p}=[e_{i,i+p-1},e_{i+p-1,i+p}],\\
&e_{i+p,i}=[e_{i+p-1,i},e_{i+p,i+p-1}],
\end{align}
\es
for any $p=1,\ldots,j-i,\ j>i$, one gets the following formulas:
\bs{gz3}
\ba
&e_{i,j}=-\frac{1}{2g}\sum_{k_i,\ldots,k_{j-1}}
\prod_{m=i}^{j-1}
\frac{\prod\limits_{\stackrel{r_{m+1}=1,}{r_{m+1}\neq k_{m+1}}}^{m+1}
(\g_{m,k_{m}}-\g_{m+1,r_{m+1}}-g)}
{\prod\limits_{s_{m}\not=k_{m}}(\g_{m,k_{m}}-\g_{m,s_{m}})}\cdot
\prod_{m=i}^{j-1}T_{\g_{m,k_m}}^{-g},\label{r2}\\
&e_{j,i}=\frac{1}{2g}\sum_{k_i,\ldots,k_{j-1}}
\prod_{m=i}^{j-1}
\frac{\prod\limits_{\stackrel{r_{m-1}=1,}{r_{m-1}\neq k_{m-1}}}^{m-1}
(\g_{m,k_{m}}-\g_{m-1,r_{m-1}}+g)}
{\prod\limits_{s_{m}\neq k_{m}}(\g_{m,k_{m}}-\g_{m,s_{m}})}\cdot
\prod_{m=i}^{j-1}T_{\g_{m,k_m}}^{g},\label{r3}
\end{align}
\es
were sum is performed  over integers $k_i,\ldots,k_{j-1}$,  such that $1\leq k_m\leq m$ for all $m=i,\ldots, j-1$.

Our next goal is to use commutation relations in Lie algebra $\gl(n,\R)$ and the results of the previous subsection to establish certain identities on the coefficients of difference operators \rf{r2} and \rf{r3}.

For any pair $(i,j)$, $i<j$ denote by $S_{i,j}$ the set  of all $(j-i)$-tuple of integers $\kk=(k_i,\ldots,k_{j-1})$, such that
$1 \leq k_m\leq m$ for all $m=i,\ldots, j-1$,
and for each  $\kk=(k_i,\ldots,k_{j-1})\in S_{i,j}$ set
\bs{r5}
\ba
&c_{i,j}^{\bold k}(\g)=
\prod_{m=i}^{j-1}
\frac{\prod\limits_{\stackrel{r_{m+1}=1,}{r_{m+1}\neq k_{m+1}}}^{m+1}
(\g_{m,k_{m}}-\g_{m+1,r_{m+1}}-g)}
{\prod\limits_{r_{m}\not=k_{m}}(\g_{m,k_{m}}-\g_{m,r_{m}})},\label{gz5a}\\
&c_{j,i}^{\bold k}(\g)=\prod_{m=i}^{j-1}
\frac{\prod\limits_{\stackrel{r_{m-1}=1,}{r_{m-1}\neq k_{m-1}}}^{m-1}
(\g_{m,k_{m}}-\g_{m-1,r_{m-1}}+g)}
{\prod\limits_{r_{m}\neq k_{m}}(\g_{m,k_{m}}-\g_{m,r_{m}})}.\label{gz5b}
\end{align}
\es
The rational functions $c_{i,j}^\jj$, $c_{j,i}^\jj$ are coefficients of difference operators presenting    $e_{i,j}$ and $e_{i,j}$ \rf{gz3}.

Next, introduce
	\beq\label{r6} \begin{split}
	a_{i,j}^{\kk}(\g)=&c_{i,j}^{\jj}(\g)\left(\T_{\jj}\right)^{-1}c_{j,i}^{\jj}(\g)\T_{\jj},\\
	b_{i,j}^{\jj}(\g)=&c_{j,i}^{\jj}(\g)\T_{\jj}c_{i,j}^{\jj}(\g)\left(\T_{\jj}\right)^{-1}	, \end{split}		\eeq
where $\T_{\jj}$ is the shift operator
\beqq
\T_{\jj}=\T_{\g_{i,k_i}}\cdots \T_{\g_{j-1},k_{j-1}}.
\eeqq
Due to \rf{r5} and \rf{r6} these function can be presented by the following products
\bs{gz6}
\ba
\hspace{-0.1cm}
	&a_{i,j}^{\jj}(\g)=
	\prod_{m=i}^{j-1}
	\frac{\prod\limits_{\stackrel{r_{m+1}=1,}{r_{m+1}\not=k_{m+1}}}^{m+1}
			\left(\g_{m,k_m}-\g_{m+1,r_{m+1}}-g \right)
		\prod\limits_{\stackrel{r_{m-1}=1,}{r_{m-1}\not=k_{m-1}}}^{m-1}
\left(\g_{m,k_m}-\g_{m-1,r_{m-1}}-g \right)}
{\prod\limits_{r_m\not=k_m}\left(\g_{m,k_m}-\g_{m,r_m}\right)
\left(\g_{m,k_m}-\g_{m,r_m}-2g\right)},\label{r7}\\ %\hspace{-0.5cm}	
&	b_{i,j}^{\jj}(\g)=\prod_{m=i}^{j-1}
\frac{\prod\limits_{\stackrel{r_{m+1}=1,}{r_{m+1}\not=k_{m+1}}}^{m+1}
\left(\g_{m,k_m}-\g_{m+1,r_{m+1}}+g \right) \prod\limits_{\stackrel{r_{m-1}=1,}{r_{m-1}\not=k_{m-1}}}^{m-1}\left(\g_{m,k_m}-\g_{m-1,r_{m-1}}+
		g \right)}
{\prod\limits_{r_m\not=k_m}\left(\g_{m,k_m}-\g_{m,r_m}\right)
\left(\g_{m,k_m}-\g_{m,r_m}+2g\right)	}.\label{r8}
\end{align}
\es
Finally, denote by $h_i(\gamma)$ the linear functions
 \beq\label{h1}
 h_i(\g)=\sum_{j=1}^i\g_{i,j}-\sum_{j=1}^{i-1}\g_{i-1,j}
 \eeq
  so that the operator $e_{i,i}$ is the operator of multiplication by $h_i(\g)/2g$.
\begin{lemma}\label{lemma2}
For any pair $(i,j) $, $1\leq i<j\leq n$ one has the identity
	\beq\label{r9}\sum_{\jj\in S_{i,j}}
\big(b_{i,j}^{\jj}(\g)-a_{i,j}^{\jj}(\g)\big)=2g \big(h_{i}(\g)-h_{j}(\g)\big).
	\eeq
\end{lemma}
{\it Proof}. This is direct corollary of the relation
\beq\label{r10}
e_{i,j}e_{j,i}-e_{j,i}e_{i,j}=e_{i,i}-e_{j,j}
\eeq
in GZ representation of  $U(\gl(n,\R))$. { Indeed, the LHS of \rf{r10} can be written as
\beq\label{gz14}
\begin{split} e_{i,j}e_{j,i}-e_{j,i}e_{i,j}=& \frac{1}{4g^2}\sum_{\jj,\jj'\in S_{i,j}}\left(c_{j,i}^{\jj'}(\g)\T_{\jj'} c_{i,j}^\jj\left(\T_{\jj}\right)^{-1}-c_{i,j}^{\jj}(\g)\left(\T_\jj\right)^{-1} c_{j,i}^{\jj'}(\g)\T_{\jj'}\right)\\
	=	& \frac{1}{4g^2}\sum_{\jj\in S_{i,j}}\left(c_{j,i}^{\jj}(\g)\T_\jj c_{i,j}^\jj\left(\T_{\jj}\right)^{-1}-c_{i,j}^{\jj}(\g)\left(\T_\jj\right)^{-1} c_{j,i}^{\jj}(\g)\T_{\jj}\right)\\
	=	& \frac{1}{4g^2}\sum_{\jj\in S_{i,j}} \left(b_{i,j}^{\jj}(\g)-a_{i,j}^{\jj}(\g)\right),	
\end{split}
\eeq
 and then \rf{r9} follows from \rf{gz0a}.
The key point here  is the fact that in \rf{gz14} the sum of the terms with $\jj\neq \jj'\in S_{i,j}$ vanishes and thus the final result does not contain the shift operators.}
\ws

\bigskip
 The next lemma is a consequence of properties of Laplace operators.
\begin{lemma}\label{lemma3}
We have the following identity of rational functions
{	\beq\label{r11}
\sum\limits_{i<j}\sum\limits_{{\jj\in S_{i,j}}}\big(a_{i,j}^{\jj}(\g)+b_{i,j}^{\jj}(\g)\big)=
\sum_{i=1}^nh_i^2(\g)-\bl^2+4g^2\brho^2.
	\eeq}
\end{lemma}
{\it Proof.} Analogous direct calculations based on the formulas (\ref{gz3}) and (\ref{h1}).
\ws

\setcounter{equation}{0}
\section{Wave function}
%\subsection{ The construction}	
 For a set of variables $\g_{i,j}$, $1\leq j\leq i\leq n$ and real positive $g$ define a kernel
 \beq \label{n1} K^{(g)}(\g)=\prod\limits_{i=1}^{n-1}\frac{\prod\limits_{j=1}^i
 \prod\limits_{k=1}^{i+1}\Gamma\left(\frac{\g_{i,j}-\g_{i+1,k}+g}{2}\right)
 	\Gamma\left(\frac{\g_{i+1,k}-\g_{i,j}+g}{2}\right)}
 {\prod\limits_{1\leq r\not=s\leq i}
 \Gamma\left(\frac{\g_{i,r}-\g_{i,s}}{2}\right)
 \Gamma\left(\frac{\g_{i,r}-\g_{i,s}+2g}{2}\right)}.
 \eeq
 Consider the wave function
 \beq\label{n2}
 \Phi_\l^{(g)}(x)=\int_{C}K^{(g)}(\g)e^{\sum_{i=1}^n h_i(\g)x_i} d\g
 \eeq
 assuming that
 \beq\label{n3} \g_{n,i}:=\l_i\in \imath\R,\qquad i=1,\ldots,n \eeq
 are fixed  parameters  and $h_i(\g)$ are defined in \rf{h1}.
 The integration contour $C$ is an imaginary plane $\imath\R^{\frac{n(n-1)}{2}}$,
 \beqq C:\ \Re \g_{i,j}=0,\qquad 1\leq j\leq i\leq n-1\eeqq
 The integral \rf{n2} absolutely converges. The proof is identical to that of \cite{GKL,IS}.
 For instance, we can use an elegant estimate \cite[(30)]{IS} by N. Iorgov and V.Shadura,
 which { states} that for fixed $ \g_{n,i}=\l_i\in \imath\R,\ \ i=1,\ldots, n$,
{\beq \label{iorgov}
|K^{(g)}(\g)|<P(\g)\exp\left(-\frac{\pi}{n}
\sum_{i=1}^{n-1}\sum_{j=1}^i|\g_{i,j}|\right)
\eeq}
 where $P(\g)$ is locally integrable function of not more than polynomial growth.

Moreover the estimate \rf{iorgov} shows that  the function  \rf{n2} is analytical on $x$ in a strip
{	\beq\label{strip}|\Im x_k|<\dfrac{\pi}{n},\qquad k=1,\ldots,n.\eeq}
 Fix  a pair $(i,j) $, $1\leq i<j\leq n$ and  a tuple $\jj=(k_i,\ldots,k_{j-1})\in S_{i,j}$.
 The following statement establishes a set of difference equations on the kernel  $K^{(g)}(\g)$ with coefficients $a_{i,j}^\jj(\g)$ and $b_{i,j}^\jj(\g)$.
 \begin{lemma}\label{lemma4}
 	We have the relation
 	\beq \label{n4}
 T_\jj\left(a_{i,j}^\jj(\g) K^{(g)}(\g)\right)=b_{i,j}^\jj(\g) K^{(g)}(\g).	
 	\eeq
 \end{lemma}
 Here $T_{\jj}$ is the shift operator with the step $2$:
 \beq\label{n5}
 T_{\jj}=T_{\g_{i,k_i}}\cdots
 T_{\g_{j-1},k_{j-1}},\qquad\text{where}\qquad T_\g f(\g)=f(\g+2).
 \eeq
 {\it Proof}. This is a direct consequence of the fundamental functional relation on the Euler
 $\Gamma$ function, $\Gamma(x+1)=x\Gamma(x)$. Indeed, in the product $K^{(g)}(\g)a_{i,j}^\jj(\g)$ we may incorporate
 all the factors { $$\frac{1}{2}\left(\g_{n,r}-\g_{m,k_m}+
 lg \right),\qquad l=0,1,2,\qquad r\not=k_n$$}
 from $a_{i,j}^\jj(\g)$ to the shifts of the corresponding $\Gamma$ functions,
 \beq\label{n6}\ts  \Gamma\left(\frac{\g_{n,r}-\g_{m,k_m}+
 	lg}{2} \right)\mapsto \Gamma\left(\frac{\g_{n,r}-\g_{m,k_m}+
 	lg}{2} +1\right) \eeq

  The application of the shift  $T_{\jj}$ returns $\Gamma$ functions \rf{n6} back to the initial position but makes a shift in other $\Gamma$ functions,
 \beqq%\label{n6}
 \ts \Gamma\left(\frac{\g_{m,k_m}-\g_{n,r}+
 	lg}{2} \right)\mapsto \Gamma\left(\frac{\g_{m,k_m}-\g_{n,r}+
 	lg}{2} +1\right) \eeqq
 which in its turn can be achieved by the multiplication of the factor
{ $$\frac{1}{2}\left(\g_{m,k_m}-\g_{n,r}+
 lg \right)\qquad l=0,1,2,\qquad r\not=k_n$$} of $b_{i,j}^\jj(\g)$. The number of $1/2$ factors is the same in both sides, so that we can cancel them.

Since the relation (\ref{n4}) plays the crucial role in our arguments, we repeat the proof by means of exact calculations.
  Using (\ref{r7}) one has, certainly
  \beq \label{l1}\begin{split}
  & T_{\jj}\left(a_{i,j}^\jj(\g)\right)=\\
  \prod_{m=i}^{j-1}&\frac{\prod\limits_{\stackrel{r_{m+1}=1,}
  {r_{m+1}\not=k_{m+1}}}^{m+1}
 	\left(\g_{m,k_m}-\g_{m+1,r_{m+1}}+2-g\right)
 \prod\limits_{\stackrel{r_{m-1}=1,}{r_{m-1}\not=k_{m-1}}}^{m-1}
 \left(\g_{m,k_m}-\g_{m-1,r_{m-1}}+2- 	g\right) }
 {\prod\limits_{s_m\not=k_m}\left(\g_{m,k_m}-\g_{m,s_m}+2\right)
 \left(\g_{m,k_m}-\g_{m,s_m}+2-2g\right)
} ,
\end{split}
\eeq
 One the other hand, the explicit calculation results to	
  \beq\label{l2}\begin{split}
  & T_{\jj}\left(K^{(g)}(\g)\right)=K^{(g)}(\g)\cdot\\	
   \prod_{m=i}^{j-1}&\prod\limits_{\stackrel{r_{m+1}=1,}{r_{m+1}\not=k_{m+1}}}^{m+1}
  \frac{\g_{m,k_m}-\g_{m+1,r_{m+1}}+g}{\g_{m,k_m}-\g_{m+1,r_{m+1}}+2-g}
  \prod\limits_{\stackrel{r_{m-1}=1,}{r_{m-1}\not=k_{m-1}}}^{m-1}
  \frac{\g_{m,k_m}-\g_{m-1,r_{m-1}}+g }{\g_{m,k_m}-\g_{m-1,r_{m-1}}+2-g}\cdot\\
  \prod_{m=i}^{j-1}&\prod\limits_{s_m\not=k_m}
  \frac{\left(\g_{m,k_m}-\g_{m,s_m}+2\right)\left(\g_{m,k_m}-\g_{m,s_m}+2-2g\right)}
  {\left(\g_{m,k_m}-\g_{m,s_m}\right)\left(\g_{m,k_m}-\g_{m,s_m}+2g\right)}.
	\end{split}
  \eeq
One can see that the product of right hand sides of \rf{l1} and \rf{l2} is precisely
$ b_{i,j}^\jj(\g)\cdot K^{(g)}(\g)$ in accordance with (\ref{r8}). \ws
 \medskip

 {\bf Remark}. {\em Note that the rational functions $a_{i,j}^\jj(\g)$ and $b_{i,j}^\jj(\g)$ themselves satisfy the difference relation with the step $2g$:
 \beq \label{n7}
 \T_\jj( a_{i,j}^\jj(\g))= b_{i,j}^\jj(\g).
 \eeq}

 Denote by
 $A_{i,j}^\jj\,(x)$ and $B_{i,j}^\jj\,(x)$ the following functions of variables $x_1,\ldots,x_N$:
 \bs{n8}
 \ba
 &	A_{i,j}^\jj\,(x)=\int_{C}a_{i,j}^\jj(\g)K^{(g)}(\g)e^{\sum_{i=1}^n h_r(\g)x_r} d\g,
 \label{n8a}\\
 &	B_{i,j}^\jj\,(x)=\int_{C}b_{i,j}^\jj(\g)K^{(g)}(\g)e^{\sum_{i=1}^n h_r(\g)x_r} d\g.
 \label{n8b}
\end{align}
\es
 Under our assumptions on the integration contour we have the following direct corollary of Lemma \ref{lemma4} :
 \begin{proposition}\label{prop1} For any  pair $(i,j) $, $1\leq i<j\leq n$ and  a tuple $\jj=(k_i,\ldots,k_{j-1})\in S_{i,j}$ { the following relations hold:}
 	\beq\label{n9}
 	A_{i,j}^\jj\,(x)= e^{2(x_i-x_j)}B_{i,j}^\jj\,(x)
 	\eeq	
 \end{proposition}
{\it Proof}. In the integral (\ref{n8a}) perform the change of variables
$\g_{a,b}\to T_\jj\g_{a,b},\,\jj\in S_{i,j}$. Then, by Lemma \ref{lemma4} and the relation
\beqq
T_{\jj} h_r(\g)= h_r(\g) +2\left(\delta_{i,r}-\delta_{j,r}\right),
\eeqq
we get the equality
\beqq 	A_{i,j}^\jj\,(x)=e^{2(x_i-x_j)}\int_{T_{\jj}C}b_{i,j}^\jj(\g)K^{(g)}(\g)
e^{\sum_{r=1}^n h_r(\g)x_r} d\g.
\eeqq
However, we can move the contour $T_{\jj}C$ back to initial position of imaginary plane since the nominator of $b_{i,j}^\jj(\g)$ kills all the poles of $K^{(g)}(\g)$ which could prevent this deformation of the contour, while the poles
 of $b_{i,j}^\jj(\g)$ are located in zeroes of  $K^{(g)}(\g)$. Thus one arrives to \rf{n9}.
  \hfill{$\Box$}
  \bigskip

 Now we are { ready} to derive the { differential} equation on $\Phi_\l^{(g)}(x)$, gathering the statements of Lemmas \ref{lemma1},
 \ref{lemma2}, \ref{lemma3}, \ref{lemma4} and Proposition \ref{prop1}.

 Set
 \bs{re1}
 \ba
 A_{i,j}\,(x)=\sum_{\jj\in S_{i,j}} A_{i,j}^\jj\,(x),\\
 B_{i,j}\,(x)=\sum_{\jj\in S_{i,j}} B_{i,j}^\jj\,(x).
 \end{align}
 \es
Since
\beq\label{r8a}
\frac{\partial}{\partial x_i}\Phi_\l^{(g)}(x)=\int_{C}h_i(\g)K^{(g)}(\g)
e^{\sum_{r=1}^n h_r(\g)x_r} d\g ,
\eeq
Lemma \ref{lemma2} implies the equality
\beq\label{r8b}
	B_{i,j}\,(x)-A_{i,j}\,(x)=2g\left(\frac{\d}{\d x_i}-
\frac{\d}{\d x_j}\right)\Phi_\l^{(g)}(x)
\eeq
 This relation together with \rf{n9} gives the system of two linear equations on $A_{i,j}\,(x)$ and $B_{i,j}\,(x)$  which solution is
 \bs{J1}
 \ba
& A_{i,j}\,(x)=2g\frac{ e^{2(x_i-x_j)}}{1- e^{2(x_i-x_j)}}
(\d_{ x_i}-\d_{ x_j})\Phi_\l^{(g)}(x),\\
& B_{i,j}\,(x)=2g\frac{ 1}{1- e^{2(x_i-x_j)}}
(\d_{ x_i}-\d_{ x_j})\Phi_\l^{(g)}(x), 	
 \end{align}
 \es
Summing up, we get the relation
 \beq \label{n11} 	
 A_{i,j}\,(x)+B_{i,j}\,(x)=-2g\cth(x_i-x_j)(\d_{x_i}-\d_{x_j})\Phi_\l^{(g)}(x).
 \eeq
 On the other hand, Lemma \ref{lemma3} together with (\ref{r8a}) results to relation
 	\beq \label{n11a}
 	\sum_{i<j}\Big(A_{i,j}(x)+B_{i,j}(x)\Big)=\Big(\sum_{i=1}^n \d^2_{x_i}-	\bl^2 +4g^2\brho^2\Big)\Phi_\l^{(g)}(x).
 \eeq
Hence, comparison of \rf{n11a} with the sum of  \rf{n11} results to the following
 \begin{proposition}\label{prop2} The function $\Phi_\l^{(g)}(x)$ satisfies the equation
 	\beq\label{n12}
 	\left(\sum_{i=1}^n\d_{x_i}^2+2g\sum_{i<j}\cth(x_i-x_j)
 (\d_{x_i}-\d_{x_j})\right)\Phi_\l^{(g)}(x)
 =\big(\bl^2-4g^2\brho^2\big)\Phi_\l^{(g)}(x)
 	\eeq\ws
\end{proposition}
It is well known that differential equation \rf{n12} is related to original Sutherland equation by means of the conjugation by the function $\prod_{p<q}|\sh(x_p-x_q)|^g$.
 Set
 \beq\label{n13}
\Psi_\l^{(g)}(x)= \prod_{p<q}|\sh(x_p-x_q)|^g\Phi_\l^{(g)}(x)=
\prod_{p<q}|\sh(x_p-x_q)|^g\int_{C}K(\g)e^{\sum_{k=1}^N h_kx_k} d\g
 \eeq
 Proposition \ref{prop2} implies
 \setcounter{theorem}{0}
 \begin{theorem} The  function $\Psi_\l^{(g)}(x)$, is the wave function of the Sutherland system,
 	\beq\begin{split}
 	H_1	\Psi_\l^{(g)}(x)=&\left(\sum_i \l_i\right)\Psi_\l^{(g)}(x),\\
 		H_2	\Psi_\l^{(g)}(x)=&-\left(\sum_i \l_i^2\right)\Psi_\l^{(g)}(x),
 		\end{split}\eeq	  	
 \end{theorem}
where
$$
H_1=\sum_{i=1}^n\frac{\d}{\d x_i},\qquad H_2=-\sum_{i=1}^n\frac{\d^2}{\d x_i^2}
+\sum_{i<j}\frac{g(g-1)}{\sh^2(x_i-x_j)}.
$$

\setcounter{equation}{0}
\section{Example}
Consider the case $n=2$ where $h_1(\g)=\g,\,h_2(\g)=\l_1+\l_2-\g$ and two basis rational functions (\ref{gz6}) are
\beq
a(\g)=(\g-\l_1-g)(\g-\l_2-g),\ \ b(\g)=(\g-\l_1+g)(\g-\l_2+g).
\eeq
According to Lemmas \ref{lemma2} and \ref{lemma3}, they satisfy the relations
\beq
\begin{split} a(\g)-b(\g)&=h_1(\g)-h_2(\g)= 2g(2\g-\l_1-\l_2),\\
a(\g)+b(\g)&=h_1^2(\g)+h_2^2(\g)-\l_1^2-\l_2^2+2g^2.
\end{split}
\eeq
Consider the wave function:
\beq\label{ac0}
\begin{split}
\Phi^{(g)}_{\l_1,\l_2}(x_1,x_2)&=\int\limits_{\imath\R}
\exp\Big\{(\l_1+\l_2-\g)x_2+\g x\Big\} \\ \times \, &
\Gamma\Big(\frac{\g-\l_1+g}{2}\Big)
\Gamma\Big(\frac{\l_1-\g+g}{2}\Big)
\Gamma\Big(\frac{\g-\l_2+g}{2}\Big)
\Gamma\Big(\frac{\l_2-\g+g}{2}\Big)
d\gamma,
\end{split}
\eeq
where $\l_1,\l_2\in\imath\R$ and $g>0$. Since
\begin{align}
T_\g a(\g)=&(\g-\l_1+2-g)(\g-\l_2+2-g),\\
T_\g K^{(g)}(\g)=&K^{(g)}(\g)\frac{(\g-\l_1+g)(\g-\l_2+g)}{(\g-\l_1+2-g)(\g-\l_2+2-g)},
\end{align}
then
\beq
T_\g\big(a(\gamma)K^{(g)}(\gamma)\big)=b(\g)K^{(g)}(\g)
\eeq
in accordance with Lemma \ref{lemma4}.

\smallskip
One can calculate integral (\ref{ac0}) in explicit terms.
Performing the shift
$\g=s+\l_1$ of the integration variable, we rewrite the  $\frak{sl}(2,\R)$--part
	 $$\phi^{(g)}_{\l_1-\l_2}(x_1-x_2)=
\exp\Big\{-\frac{1}{2}(\l_1+\l_2)(x_1+x_2)\Big\}\Phi^{(g)}_{\l_1,\l_2}(x_1,x_2)$$
of the wave function in a form
\beq\label{ex1}
\phi^{(g)}_\l(x)=e^{\frac{\l x}{2}}\int\limits_{\imath\R}
\Gamma\Big(\frac{g+s}{2}\Big)
\Gamma\Big(\frac{g-s}{2}\Big)
\Gamma\Big(\frac{g+\l+s}{2}\Big)
\Gamma\Big(\frac{g-\l-s}{2}\Big)
\exp\{s x\}d s,
\eeq
with $\l=\l_1-\l_2$, $x=x_1-x_2$. The poles of the kernel are:
\beq
\left\{
\begin{array}{l}
s_k=-g-2k,\\
s_k=-\l- g-2k,
\end{array}\right.\ \ \
\left\{
\begin{array}{l}
s_k=g+2k,\\
s_k=-\l+ g+2k,
\end{array}\right.
\eeq
where $k\in\Z_{\geq0}$. Assuming that $x>0$ and enclosing the contour of integration in the left half plain, one has
\beq
\begin{split}
\frac{1}{4\pi\imath}
\phi_\l^{(g)}(x)=e^{(\l/2-g)x}\sum_{k=0}^\infty\frac{(-1)^k}{k!}\Gamma(k+g)
\Gamma\Big(k+g-\frac{\l}{2}\Big)\Gamma\Big(-k+\frac{\l}{2}\Big)e^{-2kx}+\\+
e^{(-\l/2-g)x}\sum_{k=0}^\infty\frac{(-1)^k}{k!}\Gamma(k+g)
\Gamma\Big(k+g+\frac{\l}{2}\Big)\Gamma\Big(-k-\frac{\l}{2}\Big)e^{-2kx}.
\end{split}
\eeq
This can be written in terms of hypergeometric functions:
\beq\label{mb1}
\begin{split}
\beta^{-1}\phi_\l^{(g)}(x)=e^{(\l/2-g)x}\frac{\Gamma\Big(g-\frac{\l}{2}\Big)}
{\Gamma\Big(1-\frac{\l}{2}\Big)}F(g,g-\frac{\l}{2},1-\frac{\l}{2};e^{-2x})-\\-
e^{(-\l/2-g)x}\frac{\Gamma\Big(g+\frac{\l}{2}\Big)}
{\Gamma\Big(1+\frac{\l}{2}\Big)}F(g,g+\frac{\l}{2},1+\frac{\l}{2};e^{-2x}).
\end{split}
\eeq
where $\beta=\frac{4\pi^2\imath\Gamma(g)}{\sin(\frac{\pi\l}{2})}$. These particular hypergeometric functions are related to Legendre functions of the second kind (see \cite[3.2(45)]{BE}):
\beq\label{mb2}
\frac{e^{-\pi\imath\mu}Q^\mu_\nu(\cosh x)}{\sqrt\pi\,2^\mu\sinh^\mu x}=
e^{-(\nu+\mu+1)x}\frac{\Gamma(\nu+\mu+1)}{\Gamma(\nu+\frac{3}{2})}\;
F(\mu+\frac{1}{2},\nu+\mu+1,\nu+\frac{3}{2}\,;e^{-2x}),
\eeq
and therefore
\beq
\phi_\l^{(g)}(x)= \beta\frac{e^{\pi\imath(\frac{1}{2}-g)}}{\sqrt\pi2^{g-\frac{1}{2}}}
\sinh^{\frac{1}{2}-g} x
\Big(Q^{g-\frac{1}{2}}_{-\frac{\l}{2}-\frac{1}{2}}(\cosh x)-
Q^{g-\frac{1}{2}}_{\frac{\l}{2}-\frac{1}{2}}(\cosh x)\Big).
\eeq
Using
 the formula \cite[3.3.1 (9)]{BE} which connects Legendre functions of the first and the second kinds:
\beq\label{mb3}
\frac{e^{-\pi i\mu}}{\cos(\pi\nu)}
\Big\{Q^\mu_{-\nu-1}(z)-Q^\mu_\nu(z)\Big\}=
\Gamma(\mu+\nu+1)\Gamma(\mu-\nu)\,P^{-\mu}_\nu(z).
\eeq
we arrive to the following expression
\beq\label{mb4}
\phi^{(g)}_\l(x)=4\imath\pi^{\frac{3}{2}}2^{\frac{1}{2}-g}
\Gamma(g)\Gamma({\textstyle g-\frac{\l}{2}})
\Gamma({\textstyle g+\frac{\l}{2}})\cdot \sinh^{\frac{1}{2}-g} x\,P^{\frac{1}{2}-g}_{\frac{\l}{2}-\frac{1}{2}}(\cosh x).
\eeq
The function (\ref{mb4}) satisfies the equation
\beq
\big(\d_x^2+2g\coth x\d_x\big)\phi^{(g)}_{\l}(x)=
\Big(\frac{1}{4}\l^2-g^2\Big)\phi^{(g)}_{\l}(x).
\eeq

\section*{Acknowledgements}
The work of the first author was supported in part by RFBR and NSFB according to the research project number 19-51-18006.
The second author appreciates the support of
Russian Science Foundation, projects No. 20-41-09009, used for the proof of the statements of Sections 1 and 3. Besides, section 2 was prepared within the framework of the HSE University Basic Research Program.


\begin{thebibliography}{99}
		
\bibitem{BE} Bateman manuscript project, ed A.Erd\'elyi,
{\it Higher transcendental functions}, vol 1. McGraw-Hill, 1953.

\bibitem{Ch}  I. Cherednik,  {\it Double affine Hecke algebras}, London Mathematical Society Lecture Note Series, {\bf 319}, Cambridge University Press, Cambridge (2005).
		
\bibitem{FJM} B. Feigin, M. Jimbo, E. Mukhin, {\it Integrals of motion from quantum toroidal algebras} Journal of Physics {\bf A} Mathematical and Theoretical {\bf 50} (46) (2017) 464001.

\bibitem{HR1} M. Halln\" as, S. Ruijsenaars, {\it Joint Eigenfunctions for the Relativistic Calogero–Moser Hamiltonians of Hyperbolic Type: I. First Steps}, International Mathematics Research Notices {\bf 2014} (16) (2014) 4400-4456.
	
	
	\bibitem{HR2} M. Halln\" as, S. Ruijsenaars, {\it A recursive construction of joint eigenfunctions for the hyperbolic nonrelativistic Calogero-Moser Hamiltonians}, International Mathematics Research Notices {\bf 2015} (20) (2015) 10278-10313.



\bibitem{HR3}  M. Halln\" as, S. Ruijsenaars, {\it Joint Eigenfunctions for the Relativistic Calogero–Moser Hamiltonians of Hyperbolic Type II. The Two-and Three-Variable Cases}, International Mathematics Research Notices {\bf 2018} (14) (2018) 4404-4449.

\bibitem{H} G.J. Heckman, {\it Root systems and hypergeometric functions. II}, Compositio mathematica {\bf 64} (3) (1987) 353-373.
		
\bibitem{GT} I.M. Gelfand, M.L. Tsetlin, {\it Finite-dimensional representations of the group of unimodular mallices}, Doklady Akademii Nauk SSSR, {\bf 71} (1950) 825–828.
		
\bibitem{GKL} A. Gerasimov, S. Kharchev, D. Lebedev, {\it Representation theory and quantum inverse scattering method: the open Toda chain and the hyperbolic Sutherland model}, International Mathematics Research Notices, {\bf 2004.  17} (2004) 823-854.
			
\bibitem{IS} N. Iorgov, V. Shadura, {\it Wave functions of the Toda chain with boundary interaction} Theoretical and mathematical physics {\bf 142} (2) (2005) 289-305.
		
		\bibitem{M} I.G. Macdonald, {\it Symmetric functions and Hall polynomials} Second edition, Oxford University Press (1998).
		
		\bibitem{O} Opdam, E. M. {\it Root systems and hypergeometric functions IV}, Compositio Mathematica {\bf 67} (2) (1988) 191-209.

\bibitem{R} S. Ruijsenaars, {\it Zero-eigenvalue eigenfunctions for differences of elliptic relativistic Calogero-Moser Hamiltonians}, Theoretical and Mathematical Physics  {\bf 146} (1) (2006) 25-33.
		

		\bibitem{S}  B. Sutherland, {\it Exact results for a quantum many-body problem in one dimension. I \& II}, Physical Review {\bf A4}(5) (1971) 2019-2021 (1971), \&{\bf A5} (3) (1972) 1372-1376.
	\end{thebibliography}
\end{document}